\documentclass[aps,showpacs]{revtex4}

\usepackage{epsfig}

\begin{document}


\title{High-energy elementary amplitudes from
quenched and full QCD\thanks{To be published in Physics
Letters B} }

\author{A.F. Martini}
\email{martini@ifi.unicamp.br}
\author{M.J. Menon}
\email{menon@ifi.unicamp.br}

\affiliation{
Instituto de F\'{\i}sica ``Gleb Wataghin''\\
Universidade Estadual de Campinas, UNICAMP\\
13083-970, Campinas, SP, Brazil
}

\date{\today}

\begin{abstract}
Making use of the Stochastic Vacuum Model and the gluon gauge-invariant 
two-point correlation function, determined by numerical
simulation on the lattice in both quenched approximation and full QCD,
we calculate the elementary (quark-quark) scattering amplitudes in the 
momentum transfer space and at asymptotic energies.
Our main conclusions are the following: (1) the amplitudes decrease 
monotonically 
as the momentum transfer increases; (2) the decreasing is faster when
going from quenched approximation to full QCD (with decreasing quark masses)
and this effect is associated with the increase of the correlation
lengths; (3) dynamical fermions generate two components in the
amplitude at small momentum transfer and the transition between
them occurs at momentum transfer near 1 GeV$^2$. We also obtain 
analytical parametrizations for the elementary amplitudes, that are suitable 
for phenomenological uses, and discuss the effects of extrapolations from
the physical regions investigated in the lattice.
\end{abstract}


\maketitle

\vspace{0.5cm}

\centerline{\it To be published - Physics Letters B}

\vspace{0.5cm}

\section{Introduction}

Soft hadronic scattering, characterized by long distance phenomena, is 
one of the great challenges in high-energy physics. The difficulty arises 
from the fact that perturbative QCD can not be applied to these processes
and, presently, we do 
not know even how to calculate elastic hadron-hadron scattering 
amplitudes from a pure non-perturbative 
QCD formalism. However, progresses have been achieved through the approach 
introduced by Landshoff and Nachtmann \cite{landshoff87}, developed by 
Nachtmann \cite{nachtmann91} and connected with the
Stochastic Vacuum Model (SVM) \cite{svm}.
In that formalism the low frequencies contributions in the functional 
integral of QCD are described in terms of a stochastic process, by means of 
a cluster expansion. The most general form 
of the lowest cluster is the gauge invariant two-point field 
strength correlator \cite{svm,svm-he}

\begin{eqnarray} 
<{\bf{F}}_{\mu \nu}^{\rm C}(x){\bf{F}}_{\rho
\sigma}^{\rm D}(y)>={\delta}^{\rm CD}\frac{g^{2}<FF>}{12(N_c^2-1)}\{(
{\delta}_{\mu\rho}{\delta}_{\nu \sigma}-{\delta}_{\mu \sigma}
{\delta}_{\nu\rho}){\kappa}D({z^2/a}^{2})+ \nonumber \\
\frac{1}{2}[{\partial}_{\mu}(z_{\rho}{\delta}_{\nu
\sigma}-z_{\sigma}{\delta}_{\nu
\rho})+{\partial}_{\nu}(z_{\sigma}{\delta}_{\mu
\rho}-z_{\rho}{\delta}_{\mu \sigma})](1-{\kappa})D_{1}
({z^2/a}^{2})\}, 
\end{eqnarray} 
where $z=x-y$ is the two-point distance, $a$ is a characteristic 
correlation length, ${\kappa}$ a constant,  $g^{2}<FF>$ the gluon 
condensate and $N_c$ the number of colours 
(${\rm C}, {\rm D}=1,...,N_c^2-1$). 
The two scalar functions $D$ and $D_{1}$ describe the correlations
and they play a central role in the application 
of the SVM to high energy scattering \cite{svm-he}. Once one has information 
about $D$ and $D_{1}$, the SVM leads to the determination of the elementary 
quark-quark scattering amplitude, which constitutes important input for models
aimed to construct hadronic amplitudes. 
Numerical determinations of the above correlation 
functions, in limited interval of physical distances, exist from lattice QCD 
in  
both quenched approximation (absence of fermions) \cite{giacomo92,giacomo97} 
and full QCD (dynamical fermions included)
\cite{giacomo97full}.

In a previous paper, we determined the
elementary amplitudes from lattice QCD in the quenched approximation 
\cite{mmt}, using as framework the SVM. In this communication, we apply 
the same procedure, now taking into
account the lattice results in full QCD. Our main goal is to investigate 
the differences in the elementary amplitudes associated with quenched theory 
and
full QCD and also the effect of different bare quark masses. In addition,
we obtain analytical parametrizations for the amplitudes that are suitable 
for phenomenological uses, and discuss in some detail the region of validity
of all the results.

The manuscript is organized as follows. In Sec.
II we recall the main formulas related with the elementary amplitudes
in the context of the SVM and in Sec. III we review the parametrizations
for the correlators from numerical simulations on the lattice. 
In Sec. IV we present the results  for the elementary amplitudes
from full QCD with different quark masses and discuss the similarities
and differences with our previous result in the quenched approximation.
The conclusions and some final remarks are the contents
of Sec. V.
  
\section{Stochastic Vacuum Model}

In this Section we briefly review the main steps of the calculation scheme
that allows the determination of the elementary amplitudes through the
SVM \cite{svm,svm-he}. We refer the reader to \cite{mmt} for more details
concerning specific calculation. 

The elementary amplitude $f$ in the momentum transfer space may be expressed
in terms of the elementary profile $\gamma$ in the impact parameter space,
through the symmetrical two-dimensional Fourier transform

\begin{equation}
f(q^2)=\int_0^{\infty}bdb J_0(qb)\gamma(b) ,
\label{amplperfil}
\end{equation}
where $q^2$ is the momentum transfer, $b$ the impact parameter and $J_0$ is 
a Bessel function.

In the Nachtmann approach \cite{nachtmann91} , the study of the elementary scattering is
based on the amplitude of quarks moving on lightlike paths in an external
field, picking up an eikonal phase in traveling through the nonperturbative
QCD vacuum. In order to have gauge invariant 
Dirac's wave function solutions a Wilson loop is proposed to 
represent each quark. In this context the no-colour exchange 
parton-parton (loop-loop) amplitude can be written as 
\cite{nachtmann91}

\begin{eqnarray}
{\gamma}={\langle}Tr[{\cal{P}}e^{-ig{\int}_{loop 1}d{\sigma}_{\mu
\nu}F_{\mu \nu}(x;w)}-1]Tr[{\cal{P}}e^{-ig{\int}_{loop2}d
{\sigma}_{\rho\sigma}F_{\rho \sigma}(y;w)}-1]{\rangle}, \nonumber
\end{eqnarray}
where ${\langle}{\rangle}$ means the functional integration over the 
gluon fields (the integrations are over the respective loop areas), 
and $w$ is a common reference point from which the integrations are 
performed.

This expression is simplified in the Kr\"{a}mer and Dosch 
description by taking the Wilson loops on the light-cone. In the 
SVM the {\it leading order} contribution to the amplitude is given by 
\cite{svm-he}

\begin{equation}
 {\gamma}(b)=\eta{\epsilon}^{2}(b) ,
\label{perfilepsi}
\end{equation}
where $\eta$ is a constant depending 
on normalizations  and

\begin{eqnarray}
 {\epsilon}(b)=g^{2}{\int}{\int}d{\sigma}_{\mu
\nu}d{\sigma}_{\rho \sigma}Tr{\langle}F_{\mu \nu}(x;w)F_{\rho
\sigma}(y;w){\rangle}. \nonumber 
\end{eqnarray} 
Here ${\langle}g^2F_{\mu \nu}(x;w)F_{\rho\sigma}(y;w){\rangle}$ is 
the Minkowski version of the gluon correlator. 

After a two-dimensional integration, $\epsilon(b)$ may be 
expressed in terms of the correlation functions in 
(1) by \cite{svm-he}

\begin{eqnarray}
\epsilon(b)=\epsilon_I(b)+\epsilon_{II}(b),
\end{eqnarray}
where

\begin{eqnarray}
\epsilon_I(b)={\kappa}{\langle}g^2FF{\rangle}
{\int}_{b}^{\infty}db'(b'-b){\cal{F}}_{2}^{-1}
[D(k^2)](b'),
\end{eqnarray}

\begin{eqnarray}
\epsilon_{II}(b)=({1-\kappa}){\langle}g^2FF{\rangle}
{\cal{F}}_{2}^{-1}[\frac{d}{dk^{2}}D_{1}(k^2)](b),
\label{epsilonII}
\end{eqnarray}
For 
${\cal{D}}=D$ or $D_1$,
${\cal{D}}(k^2)={\cal{F}}_4[{\cal{D}}(z^2)]$, where
${\cal{F}}_n$ denotes a n-dimensional Fourier transform.

With the above formalism, once one has inputs for the correlation 
functions $D(z)$ and $D_1(z)$, the elementary amplitude in the 
momentum transfer space, Eq. (2),  may, in 
principle, be evaluated through Eqs. 
(3 - 6). It is important to stress that,
as constructed, this approach is intended for the high energy limit
and small momentum transfer region, namely,
$s \rightarrow \infty$, where $\sqrt s$ is the c.m. energy and 
$q^2 \lesssim {\cal O}(1)$ GeV$^2$.

\section{Lattice parametrizations}

The determination of the correlation functions through numerical simulation 
on
a lattice is made by means of the cooling technique, a procedure
that removes the effects of short-range fluctuations on large
distance correlators. The numerical results with the associated
error are usually parametrized with the functions 
\cite{giacomo97,giacomo97full}

\begin{eqnarray}
D(z) =  A_0 \exp(- \frac{|z|}{\lambda_A}) + \frac{a_0}{|z|^4}
\exp(- \frac{|z|}{\lambda_a}), 
\end{eqnarray}

\begin{eqnarray}
D_1(z)  =  A_1 \exp(- \frac{|z|}{\lambda_A}) + \frac{a_1}{|z|^4}
\exp(- \frac{|z|}{\lambda_a}),
\end{eqnarray}
where $\lambda_A$ in the non-perturbative exponential terms is the correlation 
length of the gluon field strengths. Discussions on these choices, including
the perturbative-like divergence at short distances, may be found in
Refs. \cite{giacomo97} and \cite{mmt}.

These correlation functions were first determined in the quenched SU(3) theory
and in the interval of physical distances (Euclidean space) between 0.1 
and 1 fm \cite{giacomo92,giacomo97}. After that, the effects of dynamical 
fermions 
have also been included (full QCD), for bare quark masses $a.m_q = 0.01$
and $a.m_q = 0.02$ ($a$ is the lattice spacing) and physical
distances between 0.3 and 0.9 fm \cite{giacomo97full}.
The above parametrizations are the same in all these cases and the numerical
values of the parameters, obtained from references \cite{giacomo97} and 
\cite{giacomo97full},
are displayed in Table 1.

\begin{table*}
\caption{\label{correlators}Central values of the fit parameters (without 
statistical errors) 
for the correlators (7) and (8) \cite{giacomo97,giacomo97full}.}
\begin{ruledtabular}
\begin{tabular}{|c|ccc|}
parameters & \multicolumn{2}{c}{full QCD} &  quenched \\ 
           & $m_q.a = 0.01$ & $m_q.a = 0.02$ & approximation \\
\hline
$A_0$ (fm$^{-4}$) & 14.87 & 31.04 & 128.4 \\
$a_0$             & 0.71  & 0.66 & 0.69 \\
$A_1$ (fm$^{-4}$) & 1.709 & 4.102 & 27.23 \\
$a_1$             & 0.45  & 0.39 & 0.46 \\
$\lambda_A$(fm)   & 0.34  & 0.29 & 0.22 \\ 
$\lambda_a$ (fm)  & 4.4   & 3.0 & 0.43 \\
\end{tabular}
\end{ruledtabular}
\end{table*}

\section{Results and discussion}

With the procedure described in Sec. II (see \cite{mmt} for all the 
calculational details), 
the elementary scattering amplitude in the momentum transfer space may be 
determined. The numerical results from full QCD, 
with $a.m_q = 0.01$ and $a.m_q = 0.02$, are shown in Fig. 1, together with our 
previous result in the quenched approximation. The normalized amplitudes are
displayed in the region of high momentum transfer, up to
$10$ GeV$^2$. 

\begin{figure}
\begin{center}
\includegraphics[width=8.5cm,height=6cm]{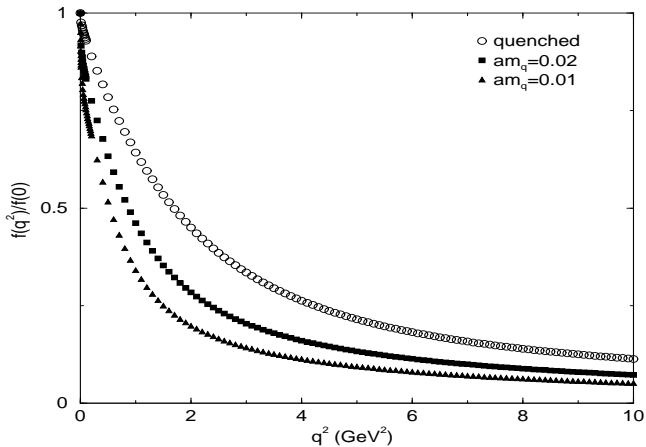}
\caption{Elementary amplitudes from full QCD and our previous result 
in quenched
approximation.}
\end{center}
\end{figure}

We see that in all the cases the amplitudes decrease smoothly as the
momentum transfer increases and they do not present any zeros in that region.
The overall basic effect of the dynamical quarks is to originate a more
rapid decrease of the amplitude, an effect that depends on the bare
quark mass: the smaller the mass the faster the decrease.
This behavior is associated with the correlation lengths
$\lambda_A$ and $\lambda_a$, since they are the only parameters that
decrease when going from full QCD (with increasing quark masses) 
to quenched approximation (see Table I).

In order to obtain analytical expressions, suitable for 
investigating distinct contributions and also for phenomenological uses, 
we have parametrized these numerical points through a sum of exponentials 
in $q^2$:

\begin{eqnarray}
\frac{f(q^2)}{f(0)}=\sum_{i=1}^n\alpha_i e^{-\beta_i q^2}.
\end{eqnarray}

By introducting a global uniform error of $0.5 \%$ in the numerical
points, we fitted the data through the program CERN-Minuit. The results 
of the fits from full QCD with $m_q . a = 0.01$ (approximation to the
chiral limit) and in the quenched approximation
are displayed in Table 2 and they are represented by the solid lines in Fig. 2,
in the regions of large and small momentum transfer. 
The corresponding exponential components in each fit
are shown in Fig. 3 for the quenched approximation and in Fig. 4 for full QCD.

\begin{table*}
\caption{\label{fitamp} Values of the fit parameters to the elementary 
amplitudes, Eq. (9), in the cases of quenched approximation and full
QCD with $m_q.a = 0.01$.}
\begin{ruledtabular}
\begin{tabular}{|c|cccc|cccc|}
Parameters: &\multicolumn{4}{c|}{$\alpha_i$}&\multicolumn{4}{c|}{$\beta_i$ (GeV$^{-2})$}\\
i =  & 1 & 2 & 3 & 4 & 1 & 2 & 3 & 4 \\
\hline
quenched     &  0.03  & -   & 0.69   & 0.28   & 55.0   & - & 0.57 & 0.09  \\
full QCD     &  0.16  & 0.54   & 0.20   & 0.10   & 170.0   & 1.52  & 0.41& 0.07\\
\end{tabular}
\end{ruledtabular}
\end{table*}

\begin{figure}
\begin{center}
\includegraphics[width=8.5cm,height=6cm]{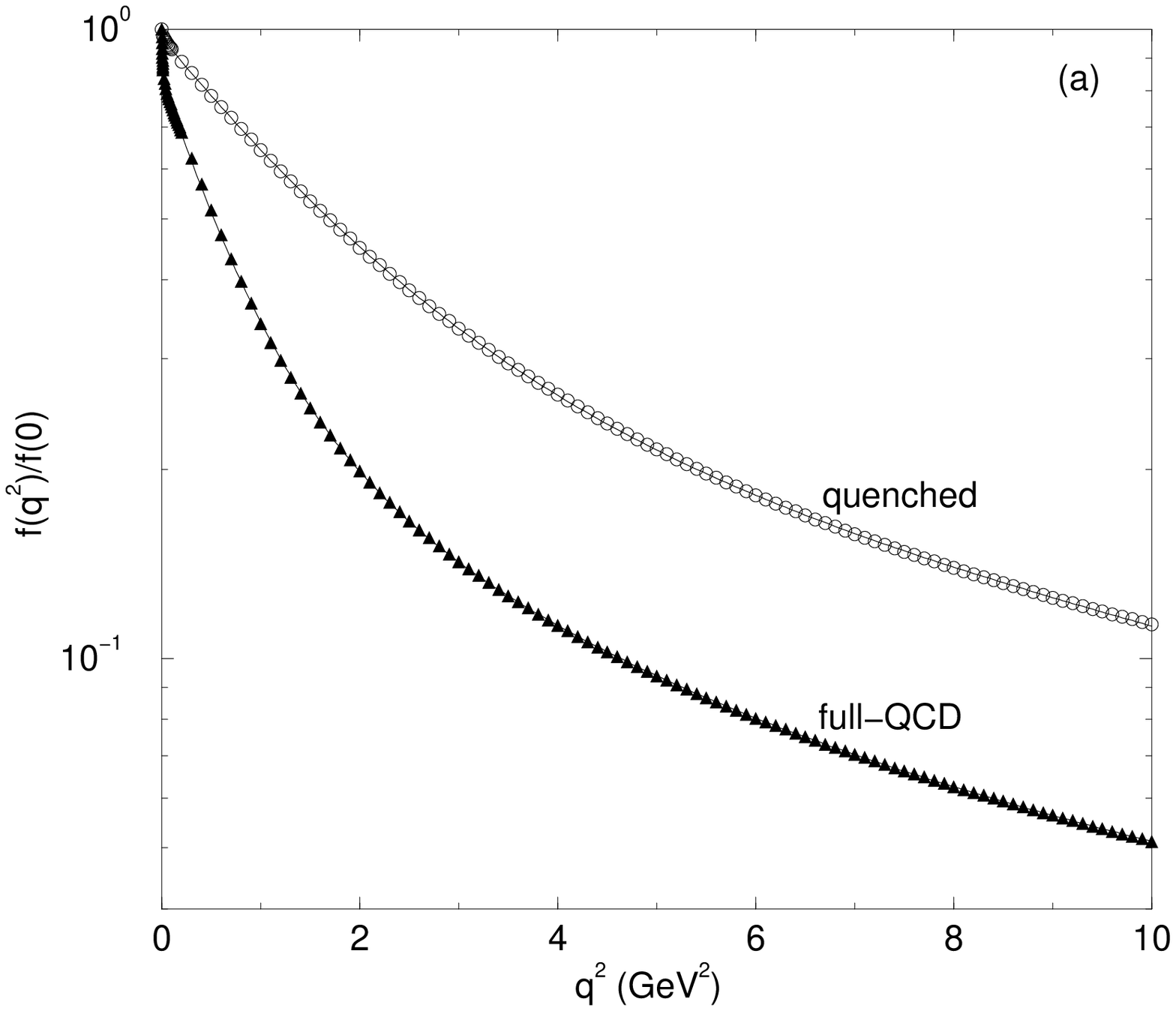}
\includegraphics[width=8.5cm,height=6cm]{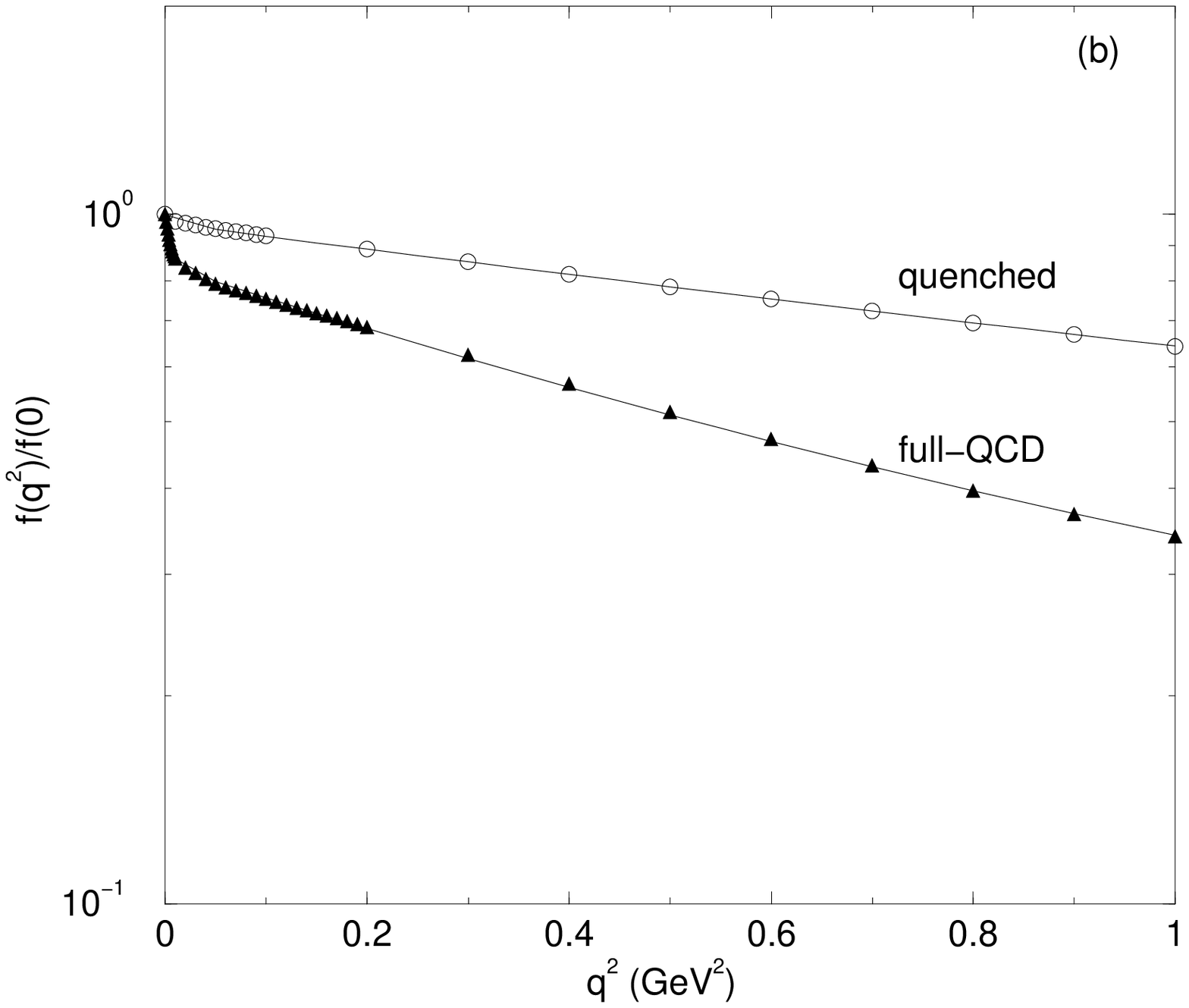}
\caption{Fits to numerical points with parametrization (9)
in the cases of full QCD for $m_q . a = 0.01$ and 
quenched approximation, in the region of large (a)
and
small (b) momentum transfer.}
\end{center}
\end{figure}

\begin{figure}
\begin{center}
\includegraphics[width=8.5cm,height=6cm]{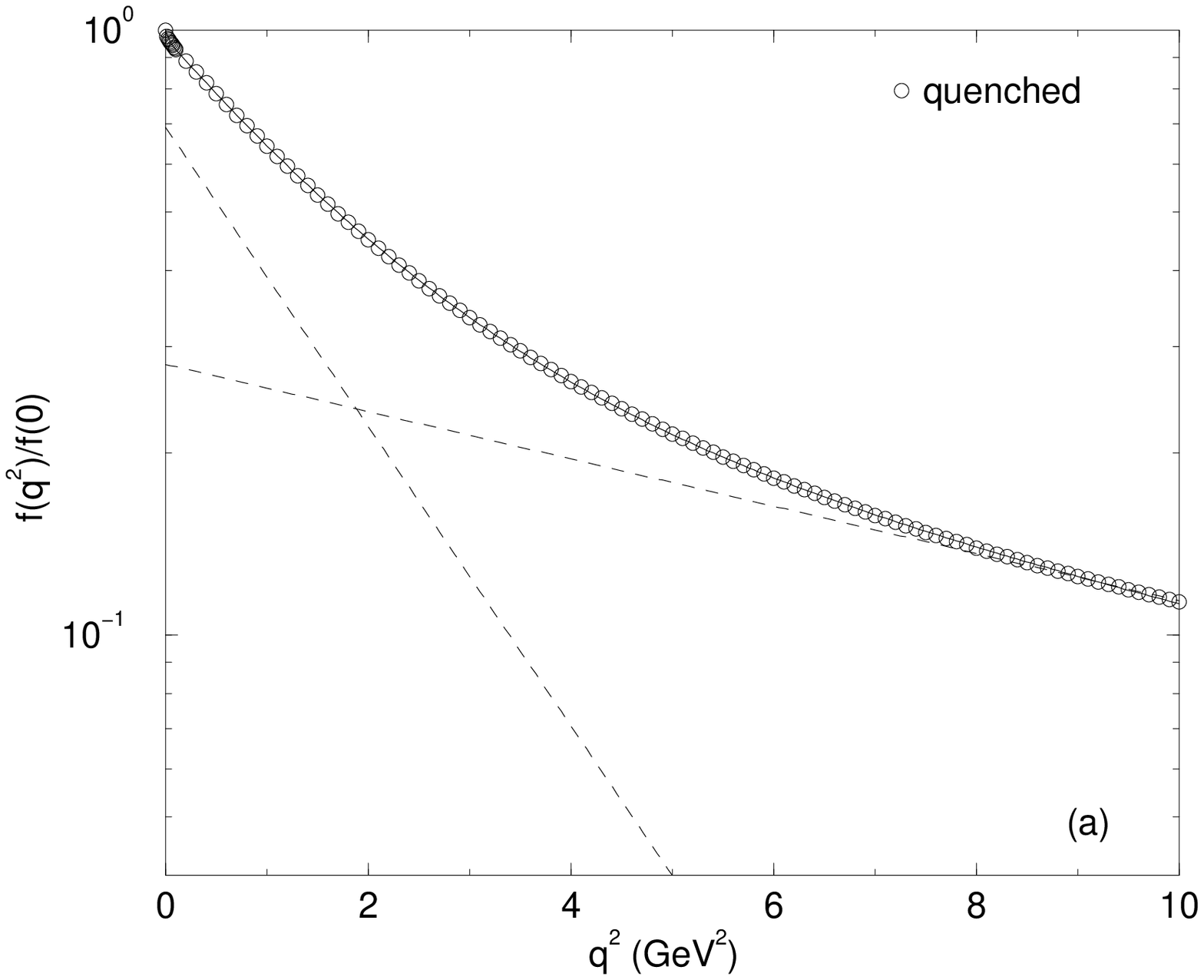}
\includegraphics[width=8.5cm,height=6cm]{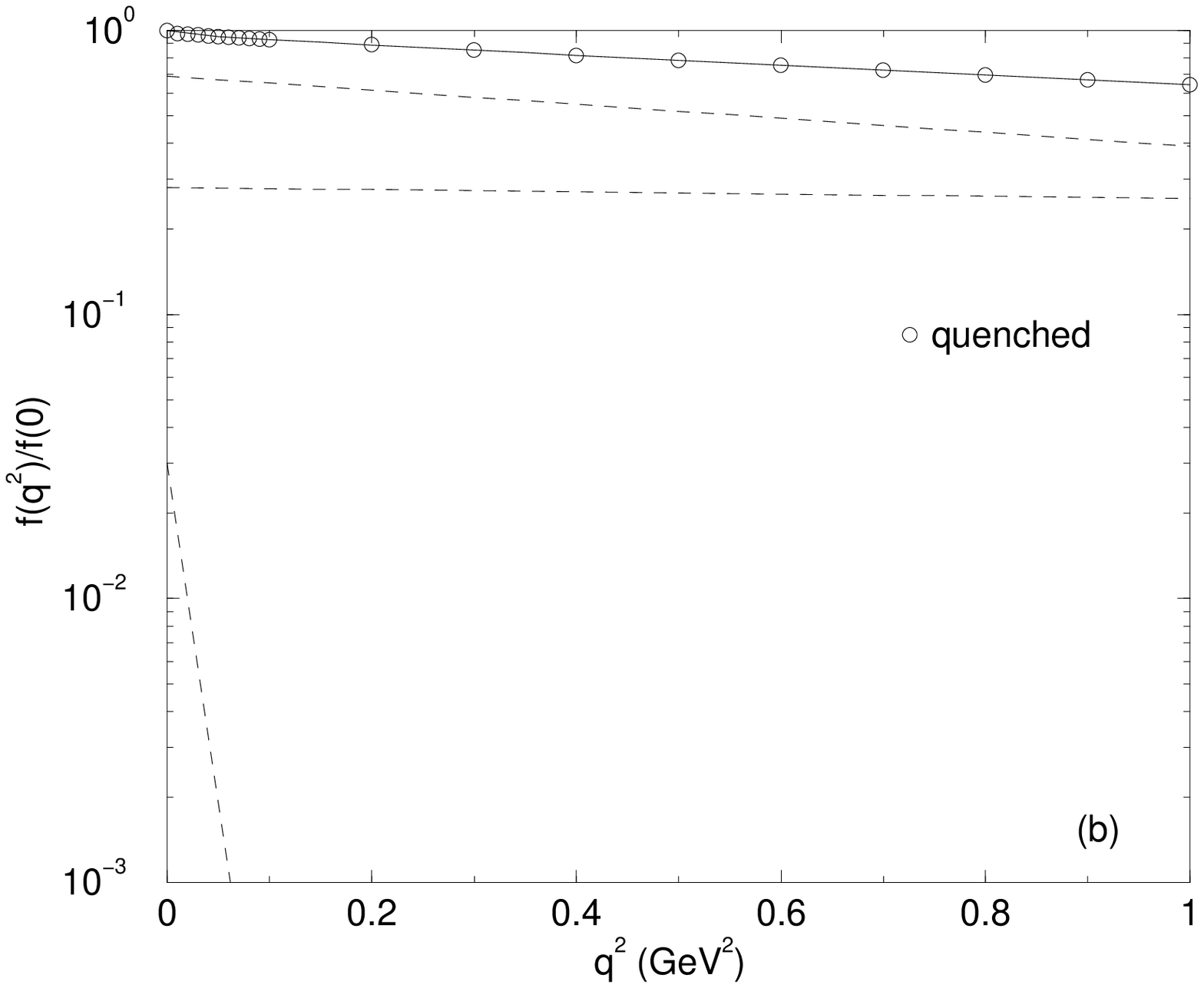}
\caption{Exponential components of the fit at large (a)
and small (b) momentum transfer in quenched approximation.}
\end{center}
\end{figure}

\begin{figure}
\begin{center}
\includegraphics[width=8.5cm,height=6cm]{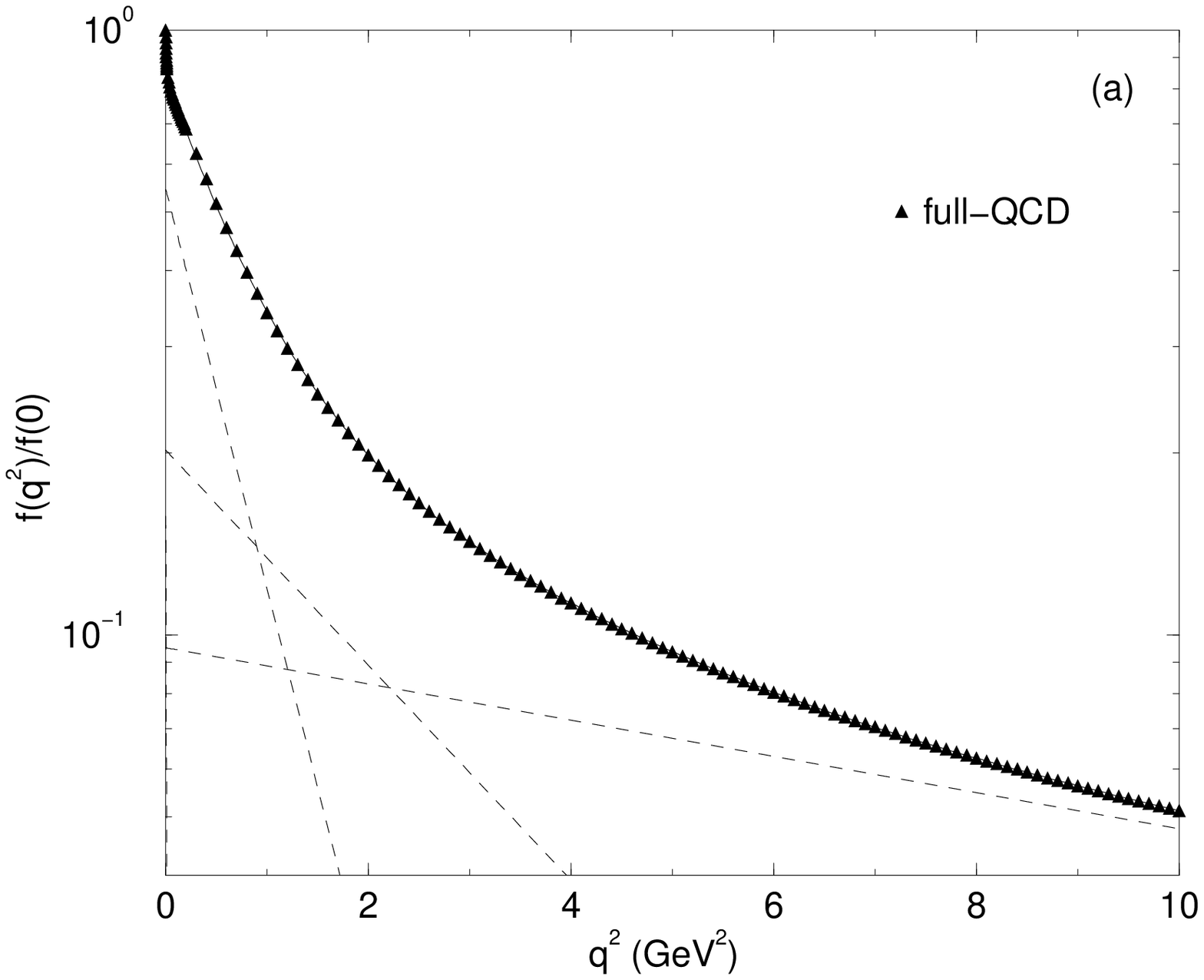}
\includegraphics[width=8.5cm,height=6cm]{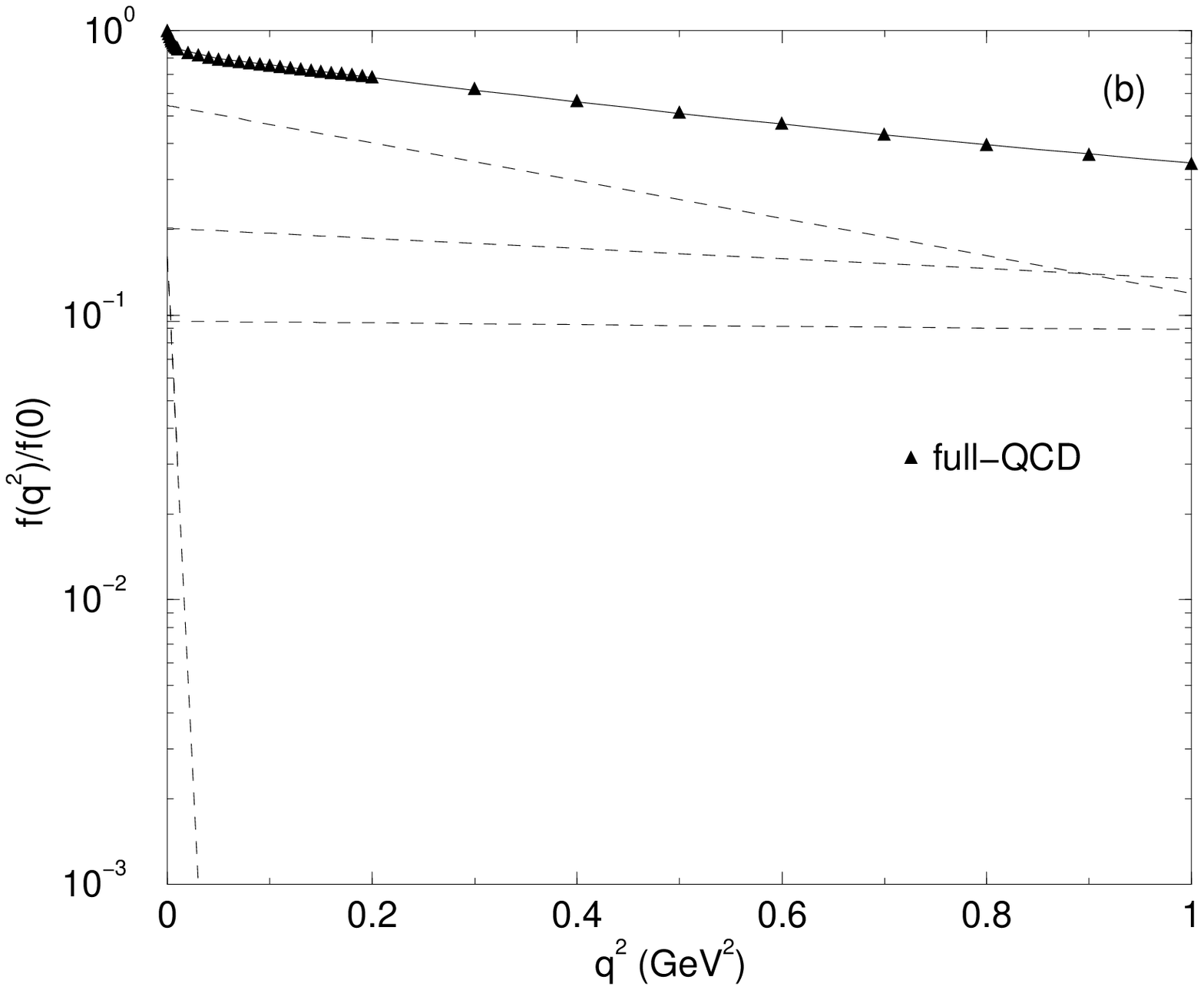}
\caption{Exponential components of the fit at large (a)
and small (b) momentum transfer in full QCD for
$m_q . a = 0.01$.}
\end{center}
\end{figure}

An immediate conclusion from these results is the presence of an additional
component in the case of full QCD. This is a central point that we
shall discuss in certain detail in what follows.

Let us start with the components with the highest slopes, which
appear in both cases in the small momentum transfer region,
namely, below  $q^2 \simeq 0.1$ GeV$^2$ (Figs. 3 and 4). As mentioned before,
the ``real" lattice results correspond to sets of discrete theoretical
points with errors, in a finite interval of physical distances, 
roughly 0.1 - 1.0 fm. Therefore, the parametrizations (7) and (8)
extrapolate this interval down and above.
The highest physical distance reached in the simulations was
0.85 fm, which corresponds to $q^2 \simeq 0.24$ GeV$^2$.
Therefore, we conclude that the components with the highest slopes
in Figs 3 and 4 ($q^2 \leq 0.1$ GeV$^2$) are associated with the
extrapolations above the physical region with ``real" lattice
results. In the same manner, the components with the smallest slopes
are connected with extrapolations down the ``real" lattice
points in physical distances and they are the responsible for the
amplitudes in high momentum region  ($q^2 \geq 6 - 7$ GeV$^2$ in
Figs. 3 and 4); therefore, they are outsite the region of validity 
of the SVM, namely, $q^2 \lesssim {\cal O}(1)$ GeV$^2$. 

We conclude that in this context only the intermediate components, 
predominant in the interval, let us say,  $0.5 \lesssim q^2 \lesssim 
2.0$ GeV$^2$, can have physical meaning
in the sense of being in agreement with the SVM  and the ``real" lattice 
results in both full QCD and quenched approximation. From Figs. 3 and 4
this ``secure" region is characterized by only one component in the case 
of the quenched approximation and two components in full QCD. The 
transition between these two components occurs
at $q^2 \simeq 1$ GeV$^2$ (Fig. 4 - left), a limit region for which the
SVM is intended for.

\section{Conclusions and Final Remarks}

In this work we have obtained analytical parametrizations for the
quark-quark scattering amplitudes in a nonperturbative QCD
framework (SVM) and using as
inputs the correlation functions, determined from numerical simulation 
on a lattice, in both quenched approximation and full QCD. The
formalism is intended for small momentum transfer ($q^2 \lesssim {\cal O}(1)$ 
GeV$^2$), asymptotic energies $s \rightarrow \infty$ and physical
distances between 0.1 and $\sim$ 0.9 fm.
As discussed in Sec. IV, these conditions put some restrictions in the
physical interpretations and, therefore, in the practical phenomenological 
uses of these amplitudes.  

However, even under the above strictly conditions we can surely extract
some novel results: (1) the amplitudes decrease smoothly as the
momentum transfer increases and they do not present zeros; (2) the decreasing 
is faster when going from quenched approximation to full QCD (with 
decreasing quark masses), and this effect is associated with the increase 
of the correlation lengths ($\lambda_A$ and $\lambda_a$);
(3) the dynamical fermions generate two contributions in the
region of small momentum transfer, which are of the same order at
$q^2 \sim$ 1 GeV$^2$ (only one contribution is present in the
case of quenched approximation).

We understand that result (3) may suggest some kind  of change in
the dynamics at the elementary level, near $q^2 \sim$ 1 GeV$^2$ and at 
asymptotic energies. If that is true, some signal could be expected
at the hadronic level.
One possibility is that this effect can be associated with the position 
of the
dip (or beginning of the ``shoulder") in the hadronic (elastic) differential
cross section data. The asymptotic condition embodied in our
result indicates that $q^2 \sim$ 1 GeV$^2$ seems in agreement with 
limit of the shrinkage of the diffraction peak, empirically verified when
the energy increases in the region 23 GeV $ \leq \sqrt s \leq $ 1.8 TeV.

At last it should be noted that if there is no new effect above
the physical distances presently investigated on the lattice ($\sim $ 0.9 fm),
the extrapolations can be considered as a good representation of the lattice
results. In this case our analytical parametrizations may be usefull
inputs for phenomenological uses in the region of small momentum
transfer and asymptotic energies.

\vspace{0.5cm}
\leftline{\bf Acknowledgments}

We are thankful to Professor A. Di Giacomo for valuable discussions 
and suggestions and to Dr. J. Montanha Neto for reading the manuscript.
This work was supported by FAPESP (Contract N. 01/08376-2 
and 00/04422-7).

\end{document}